# Gold/Silica biochips: applications to Surface Plasmon Resonance and fluorescence quenching

Thomas Mangeat[1], Alexandre Berthier[2], Céline Elie-Caille[2], Maud Perrin[2], Wilfrid Boireau[2#], Christian Pieralli[1] and Bruno Wacogne[1]

[1] Optics Department, FEMTO-ST Institute, UFR ST, 16, route de Gray, F-25030 Besançon cedex, France

[2] Clinical-Innovation Proteomic Platform, FEMTO-ST Institute, 32, avenue de l'Observatoire, F-25044 Besançon cedex, France

**Abstract:** We report Gold/Silica biochips for low cost biosensor devices. Firstly, the study of biochemical interactions on silica by means of Surface Plasmon Resonance (SPR) is presented. Secondly, Gold/Silica biochips are employed to reduce the strong quenching that occurs when a fluorophore is close to the gold surface. Furthermore, the control of the Silica-like thickness allows optimizing the distance between the metallic surface and the fluorophore in order to enhance the fluorescent signal. These results represent the first steps towards highly sensitive, specific and low cost biosensors based, for example, on Surface Plasmon Coupled Emission (SPCE) techniques.

# To whom correspondence should be sent
Tel.: +33 3 81 85 39 59 Fax: +33 3 81 85 39 98 email: wboireau@femto-st.fr

## 1. Introduction

Over the past decade optical biosensors have been largely developed to study biomolecular interactions. The most important characteristics of biosensors are specificity and sensitivity. Specificity strongly depends on the biospecific interface of biosensors. Biorecognition processes and non-specific interactions depend on the (i) surface of material, (ii) chemical functionalization and (iii) biomolecular grafting. Sensitivity depends not only on the (bio)-functionalizations but also on the biosensor architecture and transduction elements. Nowadays, techniques based on Surface Plasmon Resonance (SPR) [1] are largely used because they present a high degree of sensitivity. In this case, local change in the refractive index inside the evanescent-field region of the plasmon mode shifts the resonance angle. An angular shift of approximately $1\times10^{-4}$ degree corresponds typically to the adsorption of a biological film of about 100 pg/cm$^2$ [2, 3]. However, this technology comprises some limitations. It is well known that gold interfaces are subject to irreversible and degradative adsorption of proteins and require functionalization processes to prevent it [4]. Gold interface requires thiol-based self-assembly processes in order to build biomolecular architectures which could limit the variety of functionnalization processes. We have also to consider that a wide range of development in the field of biochip deals with the construction of molecular architectures based on silane coupling chemistries on silicon dioxide substrates. Until recently, these developments cannot take benefit of the SPR investigations. Few studies have tried to overcome this problem by chemical treating the gold surface directly or by deposing thin silicon dioxide layers. Thus a stable silica film on noble metal has been generated by applying sol-gel chemistry process [5] or through physical approaches as electron-beam evaporation [6]. Surface plasmon resonance phenomenon is extremely dependent from variations of the refractive index of the medium above the gold surface changes in a cushion of about 100 nm thickness and shift in angle is relatively limited. For instance, development of Au/SiOx chips compatible with SPR measurements requires a critical combination; namely controlling the deposition of films at a nanometer scale, their stability and reactivity while preserving the ability of the system to monitor further biomass transfer at the interface. In this way, remarkable works have been carried out recently dealing with the fabrication and characterization of stable thin film of amorphous silica deposited on SPR gold chip using the plasma-enhanced chemical vapour deposition (PECVD) technique [7-8]. In these studies, only 10 nm of silicon dioxide was deposited onto gold surface without using an adhesion layer, leading to electrochemical surface plasmon resonance works with a commercial apparatus.

Such thin layer of SiOx seems to be enough for a complete passivation of the gold layer and presents stable chemical and mechanical properties. We purpose an alternative of the previous process devoted to the conception and realization of gold/SiOx biochips compatible, in our case, with Biacore™ technology, the leader company in SPR biosensors. Afterward, the functionalization would be directly applied to simple silica substrates.

SPR is not the only method for studying biochemical interactions. A lot of optical techniques based on fluorescence [9], Fluorescence Resonance Energy Transfer (FRET) [10], or Surface Enhanced Raman Spectroscopy (SERS) [11] are widely investigated. A very large panel of solutions already provides high specificity and sensitivity (classical fluorescent probes, two photon detection, nanosphere, etc). The new challenge in fluorescence detection is to enhance the fluorescence signal and to produce highly directional signals for new highly sensitive biosensors with high signal to noise ratio (optical nanoantenna [12], Surface Plasmon Coupled Emission (SPCE) [13, 14]). However, quenching usually greatly reduces the intensity of the fluorescence signal when the bio-interaction occurs on a metallic surface. Again, gold/silica chips would be extremely interesting because they would increase the distance between the fluorescent molecule and the metallic surface of the biosensor.

In this paper, we present the fabrication of gold/silica biochips that offer an alternative to the above mentioned drawbacks concerning SPR and fluorescence sensing. We fabricated 40 nm Au biochip coated with a PECVD silica-like films of various thicknesses at low temperature. In the next part of this paper we present the process used to fabricate the gold/silica biochips as well as some characterisations of the device. In part 3, we report a new opportunity to study biological phenomena on silica layers by means of classical SPR techniques. Here, we present results concerning the fusion of vesicles on a silica surface. Another aspect is presented in part 4 of the paper. It concerns experiments conducted in order to control the quenching that occurs in the vicinity of a metallic surface. The control of the silica-like thickness allows optimizing the distance between the metal and the fluorophore. This allows suppressing the quenching. It also leads to an enhancement of the fluorescence intensity. Early experimental results show an enhancement factor of 4 compared to what is obtained on pure silica. Then, the conclusion will draw our attention to some perspective to be given to this work.

## 2. Fabrication process of gold/silica thin film.

*Fabrication process*

At first, a 2 nm thick chromium layer is deposited on a SiO$_2$ wafer (diameter: 13 mm, thickness: 0,17 mm) with plasma sputtering technology to improve the adherence of gold to the substrate. The 40 nm thick Au layer is then sputtered on the top of the Cr layer. The deposition times and the argon flow pressure have been optimized to obtain the desired thickness. Deposition times of Cr and Au layer are respectively 3 and 22 sec. Argon flow pressure and current intensity are respectively 7 µbar and 0.3 A. With these parameters, the deposition rates of the Au and Cr layers are respectively equal to 109 nm/min and 60 nm/min. Secondly, thin film au silica like layer is deposited on the top of Au layer with Plasma Enhanced Chemical Vaporization Deposition (PECVD) according the following conditions: temperature 24°C; gaz mixture SiH$_4$, N$_2$O and Arg with the respectively gaz flow 13, 51 and 35 sccm; total pressure in the reactor 0.13mBar and power 54 w at 150kHz. The deposition rate is about 660 Å/min with these experimental conditions. The thickness was controlled by the deposition duration.

*Characterization of Gold/silica thin film: roughness and SPR response.*

First of all, we tried to estimate the stoechiometry of the silica-like films. An X Photoelectron Spectroscope has been used and we deduce the following composition of the thin films: SiON$_{0.5}$C$_{0.05}$. This composition led us to call our thin films "silica-like". At present, the presence of carbon is not understood. Concerning the complex refractive index of the films, first attempts with ellipsometry are not really conclusive. However, we think that the real part of the refractive index is about 2. Up to now, we have no idea of the value of the imaginary index.

The surface roughness has been analysis with AFM techniques. The AFM used was a Nanoscope III (Veeco, Santa Barbara, CA). Imaging was performed in contact mode using NPS-oxide sharpened silicon nitride probes (Veeco). The surface topography of a 40 nm thick gold thin film and a gold/Silica thin film surface is presented in figure 1. The thickness of the silica film is approximately 10 nm. AFM images demonstrate that gold chips present a rough surface, characterized by globular gold particles with diameters of about 30 nm. Nevertheless, while the gold surface is represented by globular particles, these particles are densely packed and the section profile presents a relatively flat topography, with only 1 to 1.5 nm maximum variation in height. Moreover, the roughness calculated on 1 µm$^2$ gives value of 0.27 nm, confirming this apparent surface homogeneity. For the Gold/silica biochip, the presence of the silica thin film decreases the size of the globular particle (around 20 nm) and increases the surface roughness (0.7 nm for a 1 µm$^2$ scan).

However, a minimal surface roughness is useful to generate plasmon resonance, which is highly dependent on gold surface topography. Therefore, we have tested the SPR response of gold chips (40 nm) without silica. Experiments have been conducted with a Biacore™ 2000 system for which the value of the reflection angle is given in terms of pixels on the CCD sensor. Figure 2 shows the SPR for both commercial and home made chips. We conclude that despite a slight shift of the resonance, our gold chips exhibit a slightly better resonance quality factor than the commercial ones. The SPR response of the gold/silica biochips has been simulated and compared to the experimental records. Figure 3 shows the result. The comparison between theoretical and experimental results provide two important indications: firstly, the gold/silica SPR dip is clearly shifted compared to bare gold surfaces; secondly, the roughness increase due to silica thin film does not dramatically affect the quality factor of the SPR. In conclusion this biochip can now be used to study biochemical interaction on silica with conventional SPR equipment.

## 3. SPR lipidic interactions studied on Au/SiOx biochip

Biacore™ systems, based on SPR, are widely used to study biomolecular interactions on gold surfaces (Protein and DNA chips based on thiols chemistry as a first step of functionalization) [15]. As previously stated, the same investigation would be particularly attractive on silica to study biomolecular response on silica surfaces based on silanols chemistry. Fusion of lipid vesicles onto the functionalized silica surface has been studied by SPR. The goal is to obtain a lipidic hemi membrane conferring bio-mimetic properties to the biochip and being a versatile tool to study biomolecules/lipid interactions. These experiments are in line with works presently developed in our proteomic platform concerning new types of DNA biochips [16]. Figure 4 shows the SPR experimental results of the fusion of lipid vesicles onto a functionalized silica surface by octadecyltrichlorosilanes (OTS). On the top of the figure, the principle of lipidic vesicles fusion onto the silica surface is presented. The curve at the bottom shows a typical Biacore™ response for such an interaction. The moment when the vesicles are introduced into the microfluidic chip is clearly visible. In fact, the curve represents the shift of SPR reflection angle due to the change of the refractive index of the liquid into which the vesicles are contained. This shift is given in terms of RU (it holds for: Response Unit). After the biochemical reaction is finished (time $T_0$+4000 seconds) a buffer is used to wash the chip surface. In this way, only the lipidic layer reconstituted to OTS layer remains. At the end, a final response of 1500 RU indicates an efficient fusion of the vesicles onto the surface. This result clearly demonstrates that SPR can be used to monitor biochemical interactions on silica surfaces.

## 4. Fluorescence applications: enhancement of fluorescence signal and suppressing quenching

As mentioned above, silica thin films can be employed to reduce the strong quenching that occurs when a fluorophore is close to a metallic surface [17, 18]. Also, the control of the silica thickness allows optimizing the distance between the metallic surface and the fluorophore in order to enhance of the fluorescence signal [19]. These effects have been studied with our biochip, with Cy5 dye deposited by sedimentation on the top of gold/silica chips with several silica thicknesses.

*Experimental setup*

The experimental setup is described in Figure 5. To obtain a very good uniformity of CY5 coating, the fluorescent dye is deposited by sedimentation on the top of the gold/silica film. The same concentration of CY5 dye is deposited for several chips with the respective silica thicknesses equal to 10, 25, 50, 100 and 150nm. An Olympus IX71 fluorescence optical microscope is used to scan the silica/gold surface. The excitation is set to a mercury vapour lamp couple with a 630+20nm band pass filters. The detection is performed by means of a CCD camera (Sony-HR-XC50) with 670nm fluorescence band pass filters (CHROMA ET700/75M). Then, the average of grey level is computed from 20 images taken with each individual gold/silica chip.

*Experiments results and discussion*

The top of figure 6 shows the experimental fluorescence intensity (crosses) versus the silica thickness and theoretical modeling of quenching effect (line). The theory is a bit difficult to establish because the exact complex reflection coefficients of sputtered gold and PECVD silica are difficult to estimate. An analytical description of the quenching can be found in reference [20]. For description purposes we can say that the shape of the signal versus silica thickness is due to three main phenomena. For very small silica thicknesses, the quenching is dominant. When the silica thickness increases, a coupling between the fluorescent dipoles and the metal occurs. It is responsible for the exaltation of the fluorescence signal (together with the natural reflection of the fluorescence light onto the metallic surface). It should be noted that this enhancement very likely depends on the nanometric topology of the metal. For thicker silica layers a Fabry-Pérot behaviour can be observed. The rapid decrease of the interference pattern is due to the reduced coherence length of the fluorescence light. Then, and as it can be seen on the bottom of the figure for very thick silica layers, the influence of the metal practically disappears. The asymptotic value corresponds to what would be observed if the Cy5 dye was deposited directly onto a pure silica surface. The ratio (asymptotic value)/(maximum) defines the enhancement factor. Theoretically, it is equal to 4.3.

Looking at the experimental results, it is clear that with silica thicknesses of 10 or 25 nm, the quenching dramatically reduces the fluorescence emission. The highest fluorescence signal is achieved for a silica thickness equal to 50 nm where the enhancement factor is almost maximal. In order to confirm the fluorescence enhancement, we have compared the fluorescence obtained with a gold/silica chip (50 nm thick silica) with what is observed with a microscope slide. The result is shown in figure 7. Here, 16 pictures have been taken (horizontal axis) at various locations on the chips. The vertical axis represents the average grey levels of the images. The lower data correspond to pure silica while the upper ones to gold/silica chip. The experimental enhancement factor is equal to 3.8. Note that this can be further improved by controlling the roughness of the gold surface (work is being performed).

## 5. Conclusion.

In this paper, we have presented a new gold/silica biochip for biosensors applications. Firstly, biochemical interactions on silica were controlled by means classical SPR Biacore™ system. We demonstrated the potential of this approach through the reconstitution of lipidic membrane models devoted to, for example, the study of protein/lipidic matrix interactions. Secondly, the control of the quenching and the enhancement of the fluorescence were demonstrated using these gold/silica biochips. For a silica layer thickness of 50 nm, an enhancement factor of 3.8 was achieved.

Current work is oriented toward two different applications. The influence of the surface roughness of the gold layer on the fluorescence enhancement is currently studied. This should lead to high sensitivity biosensors. At the same time, we are looking at the possibility of using our gold/silica chips in Surface Plasmon Coupled Emission (SPCE). It was proved in literature that SPCE could increase the fluorescence signal by a factor of up to 1000. In this technique, the high directionality of fluorescence emission may greatly improve the signal to noise ratio. Furthermore, the angular position of the SPCE emission peak is strongly wavelength dependent and it is possible to use SPCE as a spectrally resolving technique. In other words, this directional detection allows using different fluorophores at a same time without the use of additional dispersive elements. The idea behind is to fabricate a biosensor that can be used to detect several biochemical reactions simultaneously.

# References

1. M. Rich R. L. and Myszka D. G., Curr. Opin. Biotechnol. **11**, 54 (2000).
2. E. Stenberg, B. Persson, H. Roos, *et al.* J. Colloid Interface Sci. **143,** 513 (1991).
3. W. D. Wilson, Science **295**, 2103 (2002).
4. J. E. Frew and H.A. Hill., Eur J Biochem. **172,** 261 (1988).
5. D. K. Kambhampati, T.A.M. Jakob, J.W. Robertson, *et al.* Langmuir, **17**, 1169 (2001).
6. H. B. Liao, W. Wena and G. K. L. Wong. J. App. Phys. **93** 4485 (2003).
7. S. Szunerits, R. Boukherroub, Langmuir **22,** 1660 (2006).
8. S. Szunerits, Y. Coffinier, S. Janel, *et al.*, Langmuir **22,** 10716 (2006).
9. W.C. Chan, D. Maxwell, X. Gao, *et al*, Curr. Opin. Biotechnol. **13**, 40 (2003).
10. I.L. Medintz, A.R. Clapp, H. Mattoussi, Nat. Mater. **2, 630** (2003).
11. D.A. Stuart, A.J. Haes, A.D. McFarland, *et al., in Proceedings .SPIE, International Society of Optical Engineering 2004*, **5327**, pp. 60-73.
12. S. Kühn, U. Håkanson, L. Rogobete, *et al.*, Phys. Rev. Lett. **97**, 017402 (2006).
13. C. D. Geddes, I. Gryczynski, J. Malicka, *et al.* J.Fluoresc. **14**, 119 (2004).
14. G. Wintera and W. L. Barnes, App.Phys.Lett. **88**, 051109 (2006).
15. V. **Mansuy-Schlick**, R. Delage-Mourroux, M. Jouvenot, *et al.* Biosensors and Bioelectronics, **21**, 1830 (2006).
16. W. Boireau, J.C. Zeeh, P.E. Puig P.E.,*et al.*, Biosensors and Bioelectronics, **20**, 1631 (2005).
17. P. Anger, P. Bharadwaj, and L. Novotny, Phys. Rev. Lett. **96**, 113002 (2006).
18. G. Schneider, G. Decher, N. Neramourg, *et al.*, Nano Lett. **6**, 530 (2006).
19. J. Zhang and J. R. Lakowicz, Opt. Express **15**, 2598 (2007).
20. T. Pagnot, D. Barchiesi, D. Van Labeke, *et al.*, Optics Letters, **22**, 120 (1997).

**Figure caption :**

Figure 1 :   Atomic Force Microscopy of the thin films. Top gold only (rms = 0.3 nm). Bottom gold/silica film (rms = 0.7 nm).

Figure 2 :   Comparison between commercial gold chips and home made chips. Despite a shift of the resonance, our chips exhibit a better resonance quality factor.

Figure 3 :   SPR response of the gold/silica chips. Top: theoretical estimation. Bottom: experimental result.

Figure 4 :   SPR control of a biochemical interaction on a silica surface. Top: schematic representation of the reaction. Bottom: Typical Biacore® response.

Figure 5 :   Experimental set up used for fluorescence experiments.

Figure 6 :   Experimental and theoretical results of fluorescence quenching and enhancement. Top: comparison between experience and modelling. Bottom: factors of importance that explain the shape of the modelling. Theoretically an enhancement factor of 4.3 can be achieved.

Figure 7 :   Experimental results of fluorescence enhancement. Upper data: gold/silica chip. Lower data: pure silica microscope slide.

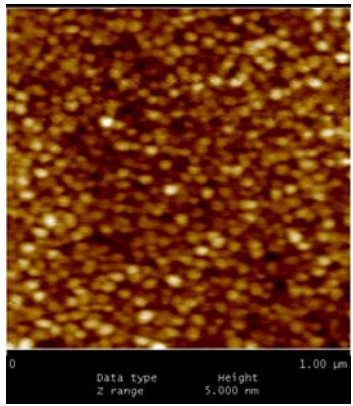
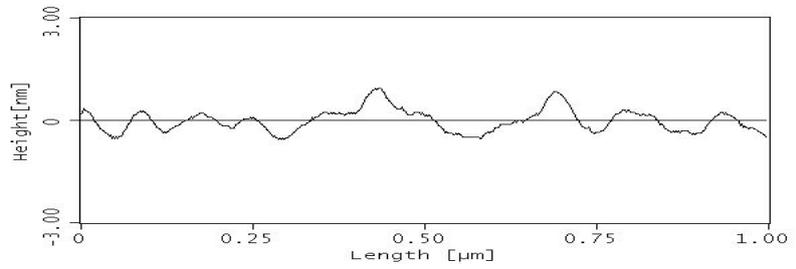

**Gold : rms = 0.3 nm**

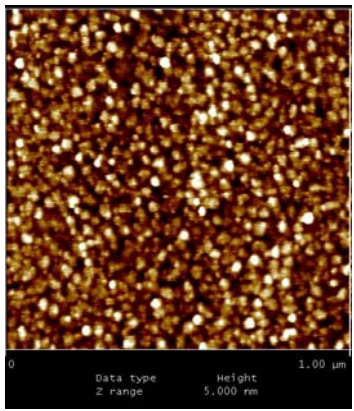
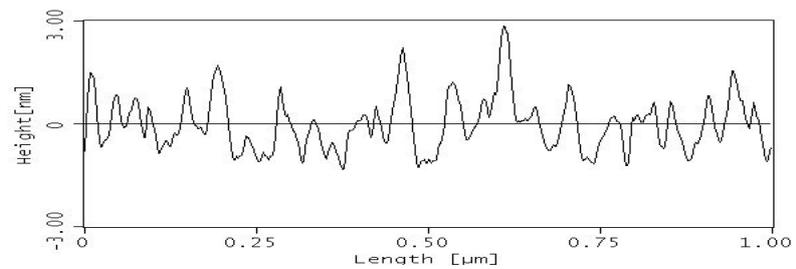

**Gold + 10 nm silica : rms = 0.7**

Figure 1

Mangeat - Gold/Silica biochips: applications to Surface Plasmon Resonance and fluorescence quenching

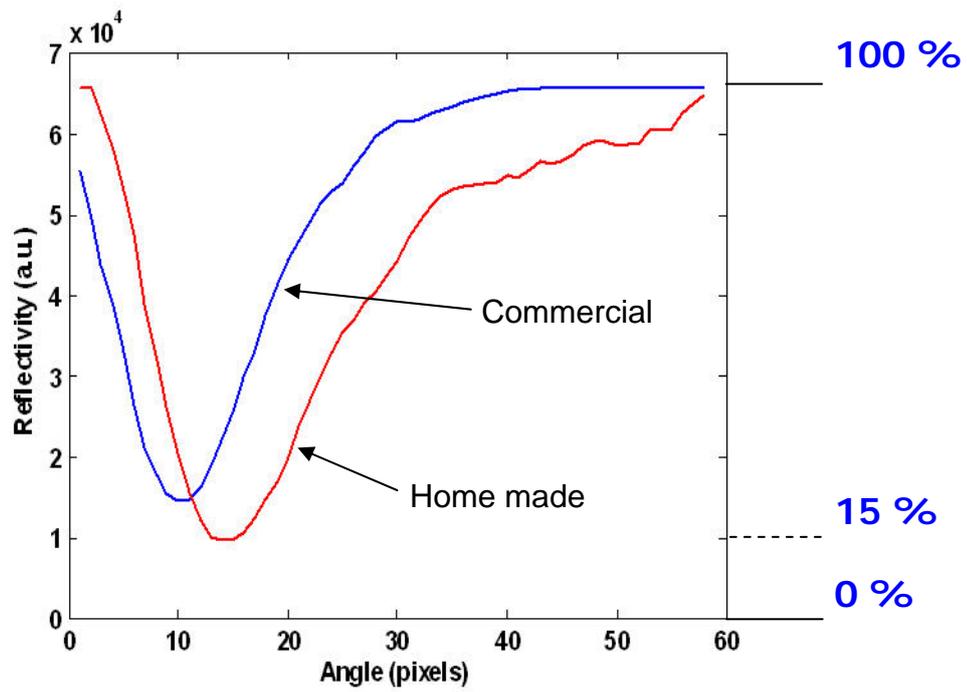

Figure 2

Mangeat - Gold/Silica biochips: applications to Surface Plasmon Resonance and fluorescence quenching

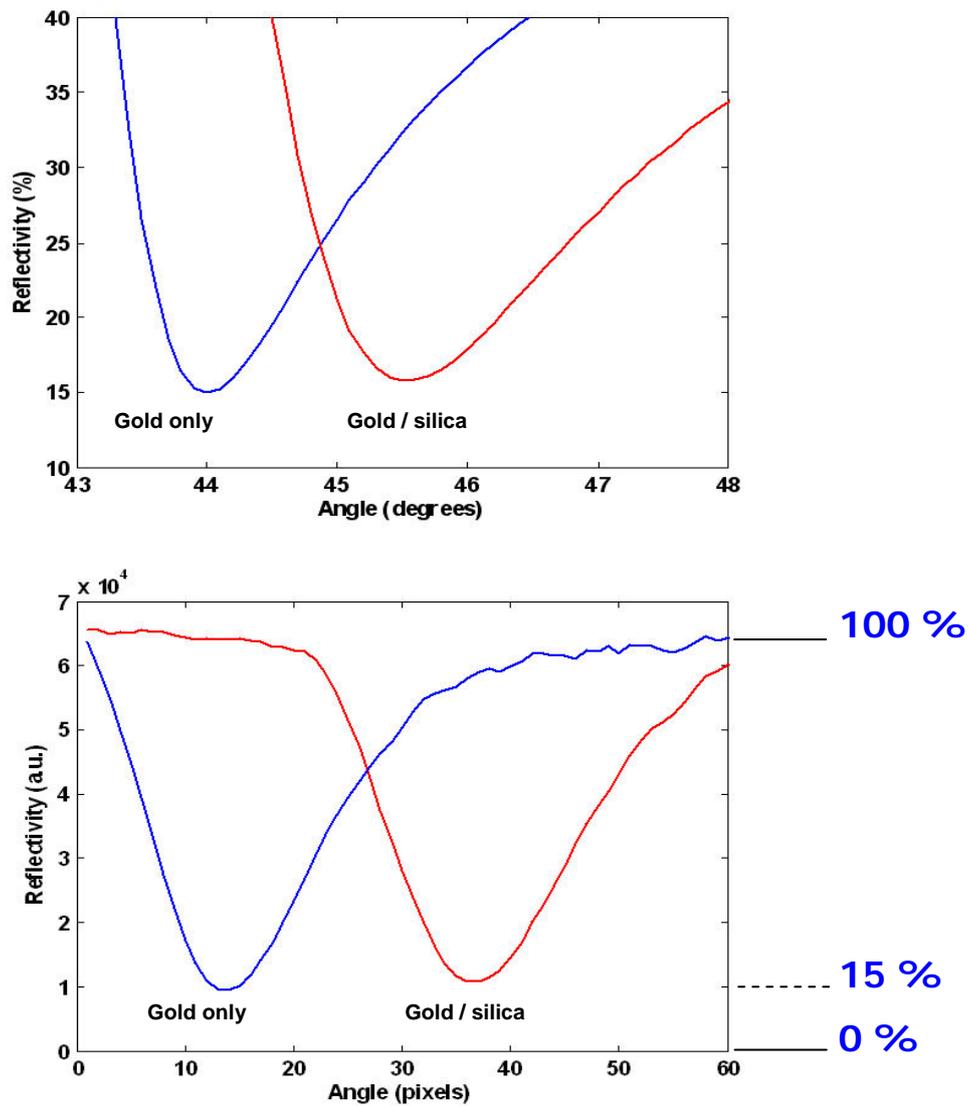

Figure 3

Mangeat - Gold/Silica biochips: applications to Surface Plasmon Resonance and fluorescence quenching

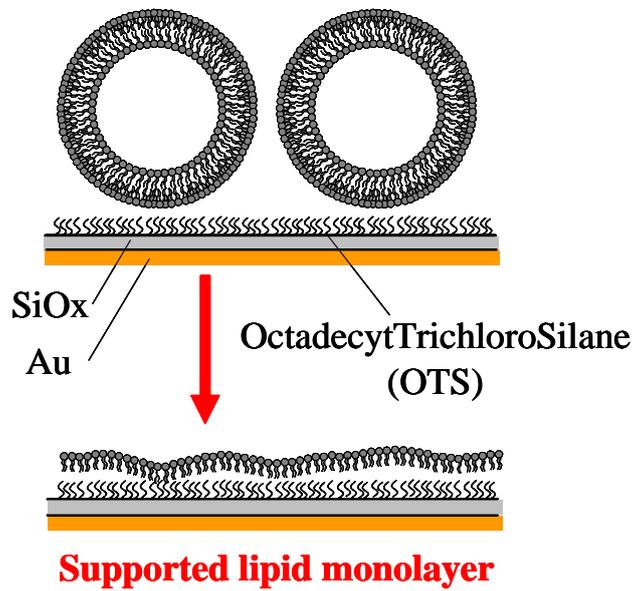

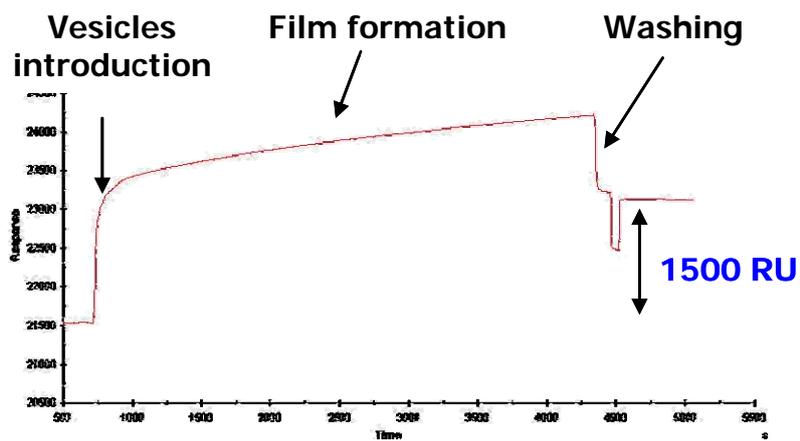

Figure 4

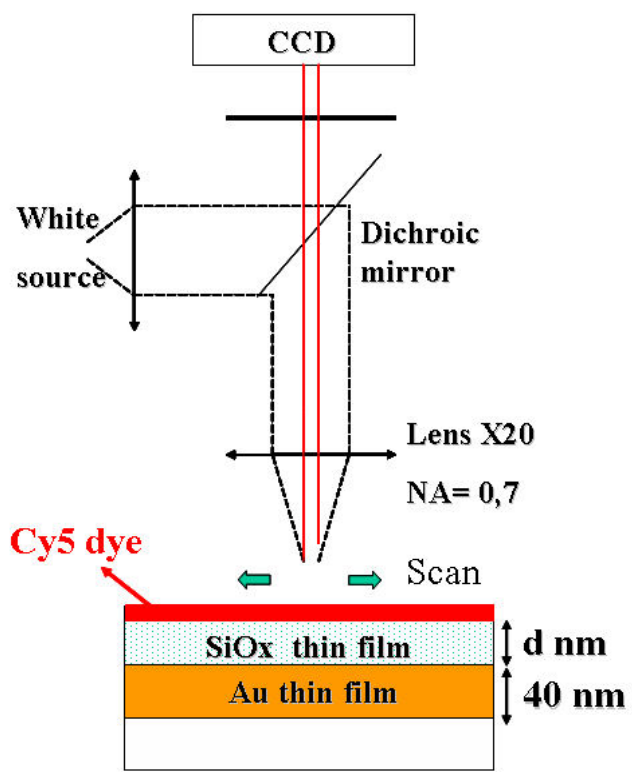

Figure 5

Mangeat - Gold/Silica biochips: applications to Surface Plasmon Resonance and fluorescence quenching

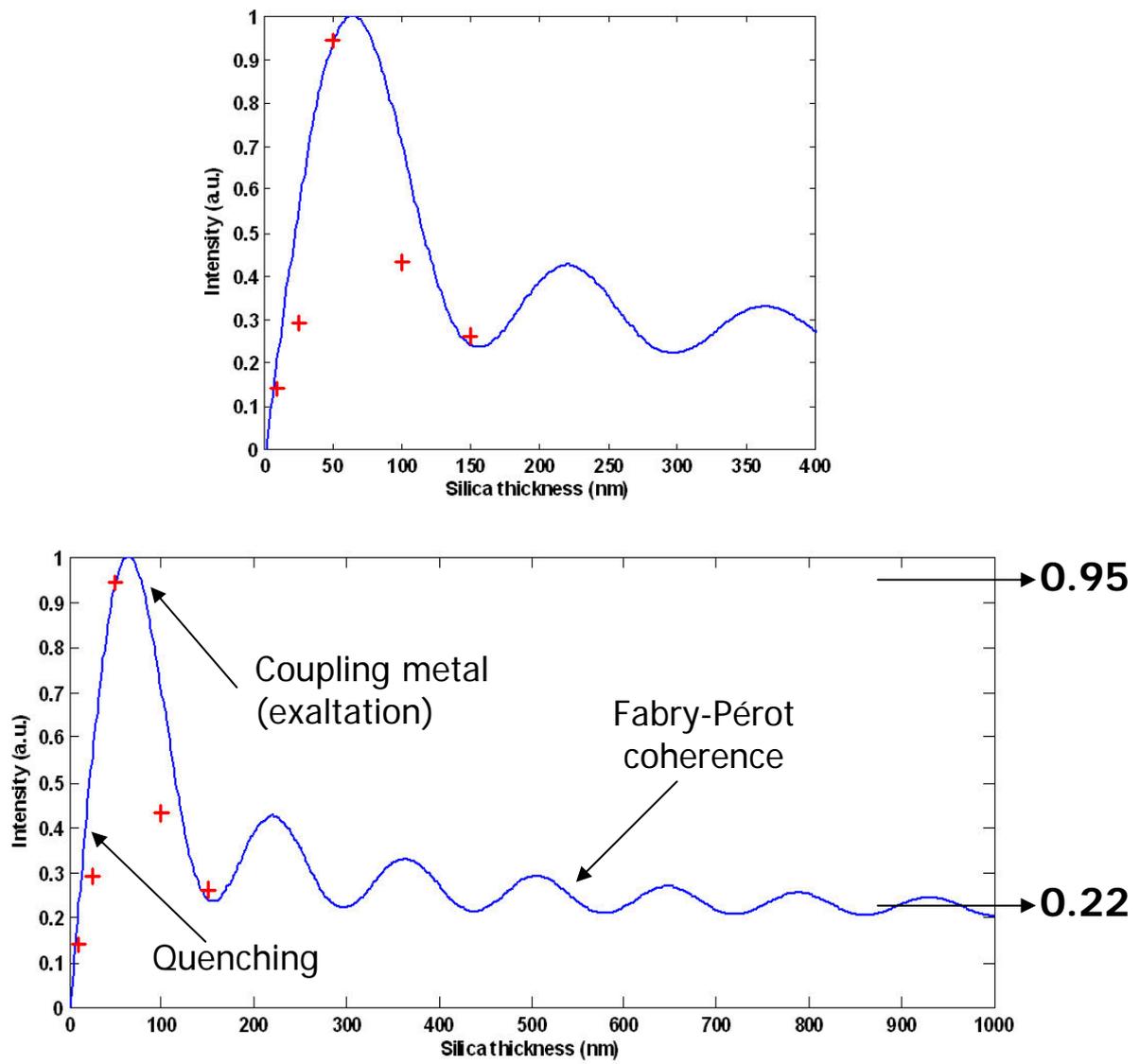

Figure 6

Mangeat - Gold/Silica biochips: applications to Surface Plasmon Resonance and fluorescence quenching

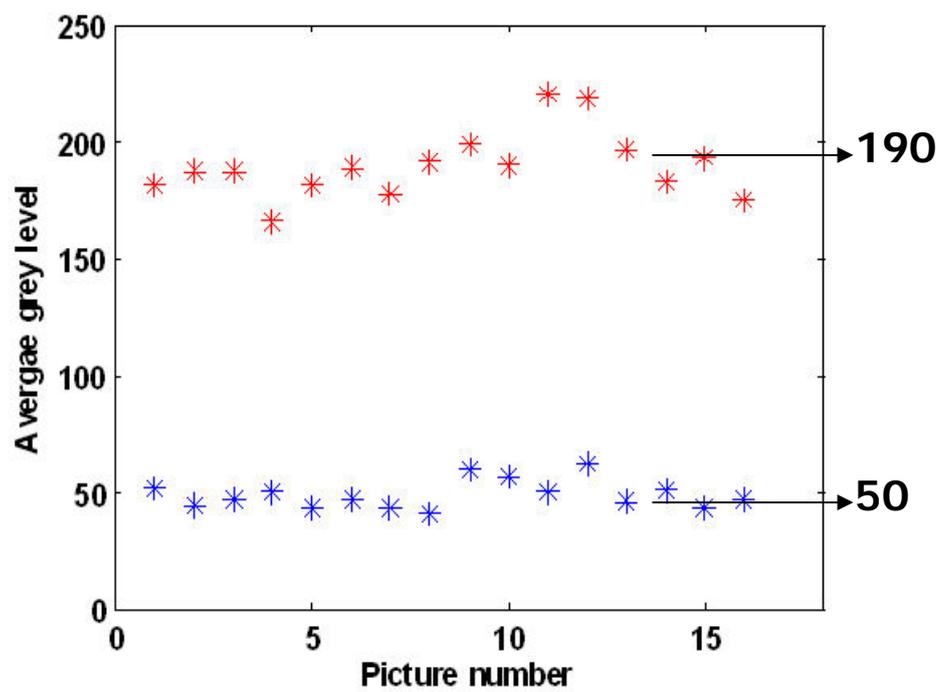

Figure 7

Mangeat - Gold/Silica biochips: applications to Surface Plasmon Resonance and fluorescence quenching